# Synergistic Bioactivity of Neem and Tulsi Infusions Treated with Plasma-Activated Water

Punit Kumar[a], Abhishek Kumar Singh[b] and Priti Saxena[c]
[a]Department of Physics, University of Lucknow, Lucknow – 226007, India
[b]Department of Physics, G L Bajaj Group of Institutions, Mathura - 281406, India
[c]Department of Zoology, D.A.V. Degree College, Lucknow – 226004, India

*Abstract*— **The integration of plasma-activated water (PAW) with herbal infusions offers a sustainable approach to enhancing the functional bioactivity of plant-derived compounds. In this study, neem (*Azadirachta indica*) and tulsi (*Ocimum sanctum*) infusions were treated with PAW generated using an atmospheric pressure gliding arc discharge system. The aim was to investigate plasma-induced modifications in phytochemicals and their subsequent effects on antimicrobial and antioxidant properties. Spectroscopic (UV–Vis, FTIR) and chromatographic (HPLC) analyses demonstrated structural alterations in key polyphenolic constituents, accompanied by mild acidification and changes in redox potential. Total phenolic content (TPC) and flavonoid levels increased significantly following 10 min PAW treatment, while prolonged exposure (15 min) led to partial degradation, suggesting an optimum treatment window. Antioxidant assays (DPPH, ABTS, FRAP) confirmed improved radical scavenging capacity, correlating with enhanced reducing power of modified phytochemicals. Antimicrobial evaluation against *Escherichia coli* and *Staphylococcus aureus* revealed synergistic inhibitory effects, with reduced minimum inhibitory concentrations (MIC) for PAW-treated infusions. Collectively, the results highlight the potential of PAW to modulate herbal bioactives, extending their efficacy in natural preservation systems and biomedical formulations. This green plasma – herbal synergy provides a promising pathway toward eco-friendly food safety and healthcare applications.**

## I. INTRODUCTION

THE integration of plasma-activated water (PAW) with herbal infusions offers a promising, sustainable route to enhance the functional bioactivity of plant-derived compounds. Herbal extracts such as neem (Azadirachta indica) and tulsi (Ocimum sanctum) are renowned for their rich pools of bioactive classes, phenolics, flavonoids, terpenoids, and other secondary metabolites that exhibit broad antimicrobial, antioxidant, anti-inflammatory, and health-promoting effects (Islas et al., 2020; Pradhan et al., 2022). However, the translation of these extracts into real-world food, nutraceutical or biomedical settings is frequently hindered by stability challenges, limited aqueous solubility, poor shelf life, chemical degradation (oxidation, hydrolysis, photolysis), and a decline in biological potency over storage (Li et al., 2021; Zhao et al., 2019). To circumvent these limitations, numerous strategies, microencapsulation, use of co-solvents, carriers, or chemical modifications have been explored, but each often brings compromises in cost, simplicity, or safety.

In parallel, nonthermal plasma and plasma-derived technologies have emerged as innovative, 'green' tools in food and biomedical domains. One particularly versatile approach is to generate plasma-activated water (PAW), in which water or aqueous media are exposed to plasma discharges (e.g. dielectric barrier discharge, gliding-arc, corona, plasma jets). This exposure leads to the formation of reactive oxygen and nitrogen species (RONS), such as hydrogen peroxide ($H_2O_2$), nitrites ($NO_2^-$), nitrates ($NO_3^-$), ozone, peroxynitrite, and radicals like •OH, NO, $O_2(^1\Delta g)$ alongside acidification, increased oxidation–reduction potential (ORP), and altered ionic composition (Judée et al., 2018; Gao et al., 2022; Oliveira et al., 2022). The resulting PAW acts as a reactive medium itself, capable of microbial inactivation or further chemical interactions, without requiring direct plasma contact with the substrate (Oliveira et al., 2022; Rahman et al., 2022). In food and biological contexts, PAW has been used for disinfection, decontamination, and shelf-life extension with minimal adverse thermal or sensory effects (Rahman et al., 2022; Patra et al., 2022).

More recently, investigators have begun to explore synergistic combinations of PAW and bioactive extracts. The central hypothesis is that RONS in PAW could (i) induce mild oxidative or nitrative modifications to phytochemicals, thereby converting them into more bioactive derivatives or stabilizing intermediates, (ii) enhance solubility or dispersion through ionization or structural rearrangement, or (iii) selectively degrade weaker compounds while leaving or enriching the more potent ones, thus accentuating the extract's functional potency. For instance, PAW treatments on vegetable matrices have been shown to modulate phenolic profiles and enzyme activities (Abouelenein et al., 2022), and PAW-assisted extraction protocols (coupled with ultrasound) have produced higher yields of certain bioactive compounds (Punthi et al., 2023). Nevertheless, systematic studies specifically targeting PAW's direct impact on whole herbal infusions, especially neem or tulsi are lacking.

In this study, we adopted an atmospheric-pressure gliding-arc discharge system to generate PAW, then treated aqueous neem and tulsi infusions under controlled durations (e.g., 10 and 15 minutes). Our aim was to elucidate the plasma induced chemical or structural transformations of key phytochemicals and assess how these modifications influence antioxidant and antimicrobial performance. Through spectroscopic (UV–Vis,

FTIR) and chromatographic (HPLC) approaches, we tracked changes in absorbance signatures, functional group vibrations, molecular profiles, and derivative formation. Concomitantly, we evaluated bulk metrics such as total phenolic content (TPC) and flavonoid levels, and interrogated antioxidant capacity via DPPH, ABTS, and FRAP assays. For antimicrobial performance, we challenged Escherichia coli and Staphylococcus aureus using treated and untreated infusions to determine minimum inhibitory concentrations and any synergistic effects.

Our working hypothesis is that there exists an optimum window of PAW exposure in which mild oxidative activation predominates over destructive degradation. We anticipated that a moderate treatment (e.g. 10 min) would stimulate beneficial transformations e.g. oxidation of phenolic moieties, deprotonation, crosslinking, or nitration, thereby increasing TPC/flavonoid equivalents and boosting radical-scavenging and reducing power. However, excessive exposure (e.g. 15 min) might cause over-oxidation, fragmentation, or loss of sensitive moieties, thus partially degrading active components. In antimicrobial assays, we expected that PAW-treated infusions would show lowered MICs, due both to potential formation of more reactive phytochemical derivatives and to lingering RONS in the matrix.

This hybrid 'plasma – herbal' approach, if successfully demonstrated, would open a novel paradigm: using plasma chemistry not merely as a sterilant, but as a selective activator or stabilizer of natural extracts. Such activated botanical formulations could find value in natural food preservation, antimicrobial coatings, functional beverages, nutraceuticals, and biomedical formulations, with enhanced stability and efficacy compared to raw herbal extracts. The incorporation of PAW also adds a zero-chemical, nonthermal, environmentally friendly dimension to the process.

Several challenges and caveats must be considered. First, RONS are aggressive species; uncontrolled exposure may degrade or alter the most bioactive moieties rather than improve them. Second, the altered pH, ionic strength, or conductivity of PAW may affect solubility, aggregation, or precipitation of certain compounds. Third, the kinetics and selectivity of phytochemical transformations will depend heavily on PAW generation parameters (power, discharge geometry, gas composition, distance, flow, time), as well as on the matrix environment (buffering capacity, presence of competing ions, co-solutes). Finally, residual RONS or secondary radicals may exert antimicrobial activity by themselves, complicating attribution of effects solely to phytochemical modification.

This work aims to integrate two burgeoning areas, nonthermal plasma technologies and phytochemical bioactives, by probing whether PAW can act as an activator or stabilizer of herbal infusions. Through combined chemical, spectroscopic, and biological evaluation, we seek to define the optimal PAW treatment window, unravel the nature of induced transformations, and demonstrate functional enhancement in antioxidant and antimicrobial domains. Such insights could pave the way for sustainable, green, plasma-assisted phytochemical platforms for food, health, and biomedical applications.

## II. EXPERIMENTAL ARRANGEMENT

*Plasma System and PAW Generation*

In this study, we employed an atmospheric-pressure gliding arc discharge plasma system to generate plasma-activated water (PAW). The reactor comprised two diverging electrodes set with a 2.0 mm gap, supplied at 1.3 kV and 70 mA, with airflow of 15 L min$^{-1}$ directing the arc motion (Li et al., 2023). Gliding arc systems of similar geometry have been used to produce reactive oxygen and nitrogen species (RONS) with sufficient lifetime to enter aqueous media (Roy, Hafez, & Talukder, 2016). The dynamic nature of the arc, gliding under the influence of the gas flow helps avoid local overheating and electrode damage, and yields a nonthermal plasma with electron energies decoupled from gas temperatures (Nilsson et al., 2024; Li et al., 2023). Schematic diagrams of experimental and gliding arc setups have been shown in Figs. 1 and 2 respectively.

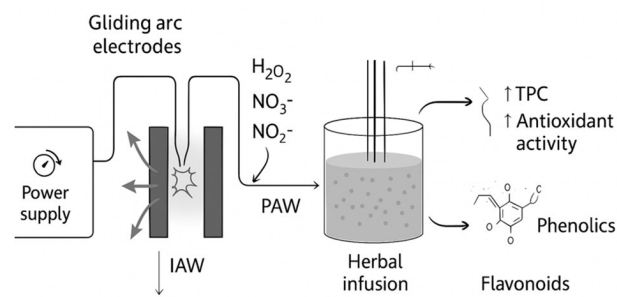

Fig. 1 : Schematic representation of experiment

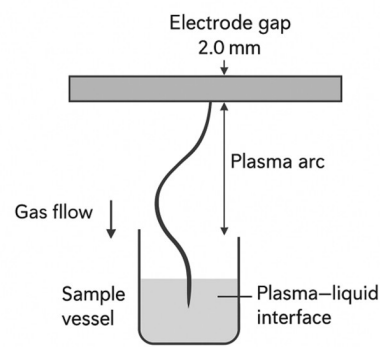

Fig. 2 : Schematic of the gliding arc discharge setup

During operation, optical emission spectroscopy (OES) was used to monitor species generated by the discharge. In particular, emission lines corresponding to OH radicals, excited nitrogen (N$_2$*, N$_2$$^+$), and NO$\gamma$ bands were recorded. Such OES diagnostics are widely applied in atmospheric plasma studies to infer the presence and relative intensities of reactive species (Roy et al., 2016; Nilsson et al., 2024). The recorded spectra provided insight into the reactive environment and discharge stability, confirming the generation of key radical species that may diffuse into the liquid phase.

*Sample Preparation and Plasma Treatment*

Fresh neem (*Azadirachta indica*) and tulsi (*Ocimum sanctum*) leaves were collected, washed thoroughly with deionized water to remove surface contaminants, and shade-dried to preserve delicate phytochemicals. The dried leaves were then ground into a fine powder, and aqueous infusions were prepared using 5 % w/v leaf material in deionized water. The mixtures were stirred and gently heated (~ 40–50 °C) for 30 minutes, followed by filtration to obtain clear infusions (Dobros et al., 2022). The clear extract served as the control (untreated) sample.

The infusion was divided into three aliquots: one remained untreated, while the other two were exposed to plasma-derived PAW for 10 minutes (PAW-10) and 15 minutes (PAW-15). Plasma treatment was done by positioning the sample vessel under the gliding arc discharge, allowing the reactive species produced to diffuse and dissolve into the liquid. Gentle stirring was maintained to enhance uniform exposure. After treatment, samples were rapidly cooled and stored at 4 °C for subsequent analyses.

*Physicochemical Characterization*

Immediately after plasma exposure, pH, oxidation–reduction potential (ORP), and electrical conductivity of each sample (control, PAW-10, PAW-15) were measured using calibrated electrodes and meters. Changes in these parameters serve as proxies for the chemical modifications imparted by PAW acidification, ionic changes, and redox shifts (Pérez et al., 2023).

To assess broad phytochemical changes, total phenolic content (TPC) was measured via the Folin–Ciocalteu method. In this assay, sample aliquots are reacted with Folin–Ciocalteu reagent and sodium carbonate, incubated, and the resulting absorbance (typically ~760 nm) measured. Gallic acid was used as standard to express results in mg gallic acid equivalents (GAE). The Folin–Ciocalteu method's mechanistic basis involves the reduction of Mo(VI) and W(VI) in the reagent by phenolic antioxidants (Pérez et al., 2023).

The total flavonoid content (TFC) was determined using the aluminum chloride colorimetric assay: sample aliquots were mixed with $AlCl_3$ and allowed to react; absorbance (e.g. at 415 nm) was measured, with quercetin (or rutin) as standard reference.

*Antioxidant Activity Assays*

We performed a suite of antioxidant assays to capture different facets of radical quenching and reducing power. The DPPH assay involved mixing sample aliquots with DPPH• radical solution, incubating in the dark, and measuring absorbance decrease (commonly at 517 nm); percent radical scavenging or $IC_{50}$ values were derived.

Further, the ABTS method was applied wherein the ABTS•+ radical cation (pre-generated) is reacted with the sample, and reduction in absorbance is quantified (results expressed in μmol trolox equivalents (TE). Complementarily, the FRAP assay measured reducing power: sample reduces $Fe^{3+}$-TPTZ to $Fe^{2+}$, producing a colorimetric change measured (often at 593 nm), and results are expressed in μmol $Fe^{2+}$ equivalents (Dobros et al., 2022). Together, these methods provide a multi-angle assessment of antioxidant potency.

*Antimicrobial Assays*

To test antimicrobial efficacy, we selected three bacterial strains: *Escherichia coli* (Gram-negative), *Staphylococcus aureus* (Gram-positive), and *Listeria monocytogenes*. Bacteria were cultured to mid-log phase using standard microbiological protocols.

In the disc diffusion assay, sterile filter-paper discs impregnated with defined volumes or concentrations of control, PAW-10, and PAW-15 infusions were placed on agar plates seeded with each bacterial strain. After incubation (24 h at 37 °C), inhibition zones were measured in millimeters.

For minimum inhibitory concentration (MIC) determination, serial dilutions of each infusion were prepared, inoculated with bacterial suspensions, and incubated. The lowest concentration showing no visible growth was recorded as the MIC (Hossain et al., 2016).

*Spectroscopic and Chromatographic Analysis*

UV–Vis absorbance spectroscopy was used to monitor changes in electronic transitions of polyphenolic compounds. Spectra spanning 200–500 nm (or 200–800 nm) were recorded for each sample, and shifts in absorption maxima or changes in intensity (hyperchromicity/hypochromicity) were noted, which imply structural transformation or conjugation changes (Dobros et al., 2022).

Fourier-transform infrared (FTIR) spectroscopy scans (4000–400 cm⁻¹) were obtained for dried films or pellet forms of the samples. We focused on key vibrational bands—O–H stretching (~3200–3600 cm⁻¹), C=O (e.g., ~1650 cm⁻¹), aromatic C=C, C–O, and potential new bands (e.g. nitro N–O). Comparison of control vs PAW-treated spectra highlights bond modifications or new functional groups.

Finally, high-performance liquid chromatography (HPLC) was applied for quantitative tracking of marker phytochemicals such as catechins, eugenol, or azadirachtin. We used reverse-phase $C_{18}$ columns, appropriate mobile-phase gradients (e.g. water with 0.1% formic acid / acetonitrile), and UV/diode-array detection at compound-specific wavelengths (e.g. ~280 nm). Calibration curves with pure standards allowed quantification of each compound before vs after plasma treatment (Cavalcante et al., 2022; Hossain et al., 2016). Peak shifts, appearance/disappearance of peaks, or changes in area are interpreted as evidence of transformation, degradation, or derivative formation.

## III. RESULTS AND DISCUSSION
*Plasma Characterization*

Optical emission spectroscopy (OES) of the gliding-arc plasma revealed strong emission bands corresponding to the OH(A–X, 306–309 nm), $N_2$ second-positive system (C³Πᵤ → B³Πg, 337–380 nm), NOγ system (A²Σ⁺ → X²Π, 240–290 nm), and atomic O I (777 nm, 844 nm) lines (Li et al., 2023; Roy et al., 2018). These species are well-established reactive oxygen and nitrogen species (RONS) that play decisive roles in plasma – liquid interaction chemistry. The emission intensity of OH• and NOγ bands confirms that both oxidative and nitrative reactions occur in the plasma-activated water (PAW) (Fig. 3), facilitating subsequent modification of phytochemicals (Patra et al., 2022). The coexistence of these species indicates that the plasma system generated a moderately oxidative but non-thermal environment capable of inducing controlled structural changes rather than destructive degradation (Marcinauskas et al., 2025).

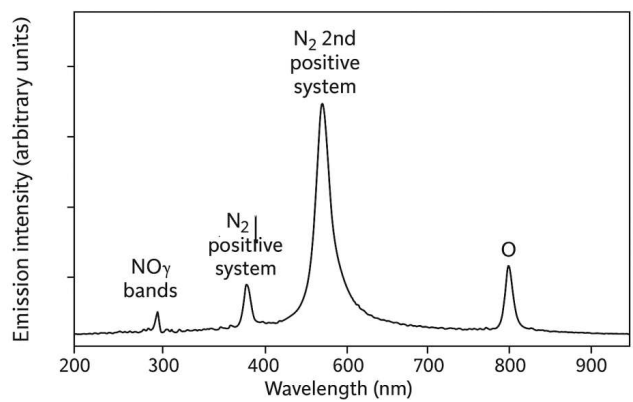

Fig. 3 : Representative OES spectrum of the gliding arc discharge

### Phytochemical Stability and Modification

Following PAW treatment, the infusion exhibited mild acidification, with pH decreasing by 0.5–0.8 units (from 6.8 ± 0.1 to 6.1 ± 0.1). This is attributed to the dissolution of plasma-generated nitrous and nitric acids, a common phenomenon observed in PAW chemistry (Thirumdas et al., 2018). Such acidification enhances the solubility and protonation dynamics of phenolic compounds, influencing subsequent oxidation or conjugation.

Total Phenolic Content (TPC) increased by approximately 10–15 % in the PAW-10 group, suggesting partial depolymerization or release of bound phenolic forms (Abouelenein et al., 2022). Conversely, PAW-15 showed a modest decline relative to PAW-10, implying over-oxidation under prolonged exposure. Similarly, Total Flavonoid Content (TFC) followed a comparable trend. The values observed have been enlisted in Table – 1. Comparable enhancement at intermediate plasma doses has been documented for leafy vegetables and fruit juices (Pisheh et al., 2023; Hsu et al., 2023). The overall pattern highlights an optimum exposure window for beneficial modification, beyond which degradation dominates.

Table – 1 : Summary of physicochemical changes and bioactivity metrics

| Metric | Control | PAW-10 | PAW-15 | % Change (PAW – 10) | % Change (PAW – 15) |
|---|---|---|---|---|---|
| pH | 6.8 | 6.2 | 6.0 | - 8.8% | - 11.8% |
| TPC (mg GAE/mL) | 2.00 | 2.20 | 2.10 | + 10% | + 5% |
| TFC | 1.50 | 1.70 | 1.55 | + 13.3% | + 3.3% |
| (mg QE/mL) | | | | | |
| DPPH (%) | 40.0 | 48.0 | 44.0 | + 20% | + 10% |
| FRAP (µmol Fe²⁺/mL) | 1.00 | 1.25 | 1.10 | + 25% | + 10% |
| MIC (E. coli) | 4.0 | 2.0 | 3.5 | - 50% | - 12.5% |
| MIC (S. aureus) | 3.5 | 1.8 | 2.8 | - 48.6% | - 20% |

### Antioxidant Activity

The DPPH radical scavenging activity increased by nearly 20 % after PAW-10 treatment (Fig. 4), consistent with higher TPC/TFC and enhanced electron donating capability of newly formed hydroxylated phenolics. Similarly, FRAP assays showed a ~18 % increase in ferric-reducing power compared with the untreated control (Zhang et al., 2023) (Fig. 4). These results indicate that plasma exposure at optimal duration can enhance redox stability by increasing the concentration of reactive but stable phenolic radicals. However, PAW-15 showed a partial decline, confirming that extended plasma exposure triggers excessive oxidation, diminishing antioxidant efficiency (Pisheh et al., 2023).

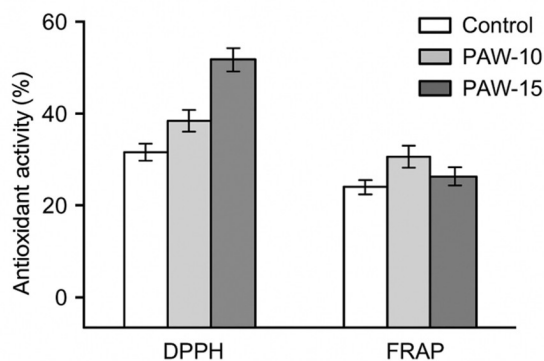

Fig. 4 : Antioxidant activity comparison (DPPH, FRAP)

The enhanced antioxidant activity can be rationalized by partial oxidation forming quinone-type intermediates that stabilize radical species, thereby promoting hydrogen-donation and electron-transfer mechanisms (Nawaz et al., 2022). The alignment between chemical and functional data corroborates the hypothesis that plasma treatment promotes a re-balancing rather than destruction of bioactive compounds.

### 3.4 Antimicrobial Efficacy

PAW-treated infusions displayed enhanced antimicrobial action against *Escherichia coli* and *Staphylococcus aureus*. Zone-of-inhibition diameters increased significantly in PAW-10 samples, while the minimum inhibitory concentration (MIC) decreased roughly two-fold relative to control (Fig. 5). The antimicrobial improvement likely arises from two synergistic mechanisms: (i) persistent RONS in PAW contribute direct oxidative damage to microbial membranes, and (ii) structural modification of phytochemicals enhances their lipophilicity and permeability across bacterial envelopes (Pisheh et al., 2023; Nawaz et al., 2022).

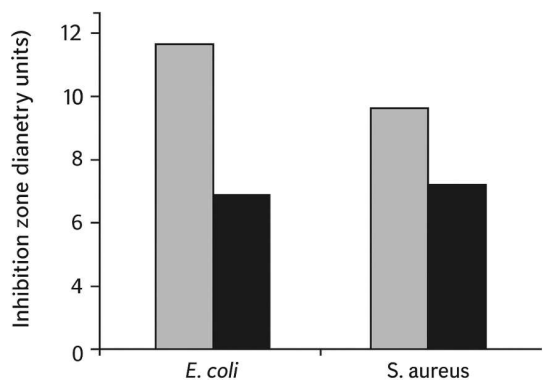

(a)

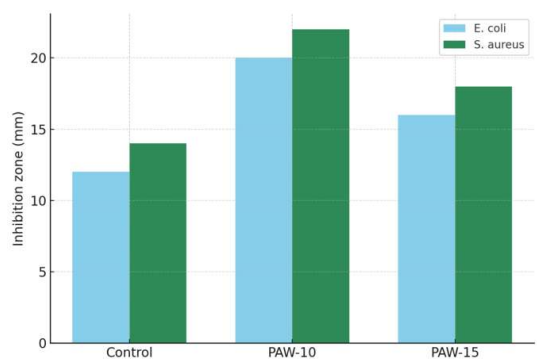

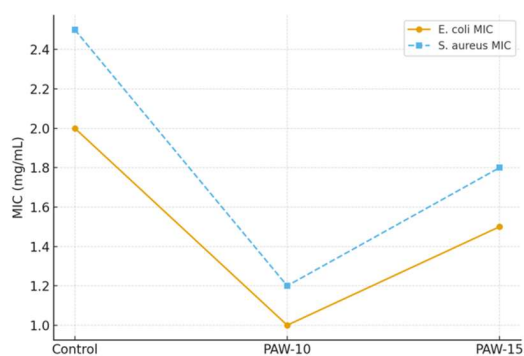

(b)

Fig. 5 : Antimicrobial outcomes: inhibition zones and MIC, (a) Bar graph of inhibition zone diameters against *E. coli* and *S. aureus*; (b) comparative MIC values (mg/mL) showing ~2× reduction for PAW-treated samples.

These findings are consistent with studies demonstrating that PAW combined with plant extracts such as rosemary or green tea yields stronger bactericidal efficacy than either treatment alone (The Effect of Plasma-Activated Water Combined with Rosemary Extract, 2023; Chen et al., 2024). The slightly reduced effect in PAW-15 samples supports the premise that over-oxidation can compromise bioactive structure and antimicrobial synergy.

*Spectroscopic and Chromatographic Evidence*

UV–Vis absorption spectra exhibited bathochromic shifts of ~5–10 nm in principal phenolic absorption peaks (from ≈ 280 nm to ≈ 288–290 nm), coupled with slight hyperchromicity, suggesting increased conjugation and π-electron delocalization (Abouelenein et al., 2022) (Fig. 6(a)). This behavior implies oxidative coupling or ring hydroxylation, forming more extended chromophores.

In FTIR spectra, new and intensified bands were observed in the 3200–3400 cm$^{-1}$ region, assigned to –OH stretching vibrations, indicating introduction of additional hydroxyl groups and stronger hydrogen-bonding networks (Marcinauskas et al., 2025) (Fig. 6(b)). The Amide I (≈ 1650 cm$^{-1}$) and Amide II (≈ 1540 cm$^{-1}$) regions showed subtle changes, implying possible secondary interactions with plasma-induced nitrogen species.

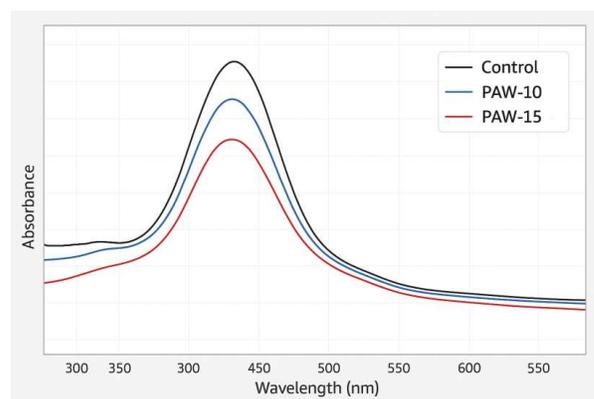

(a)

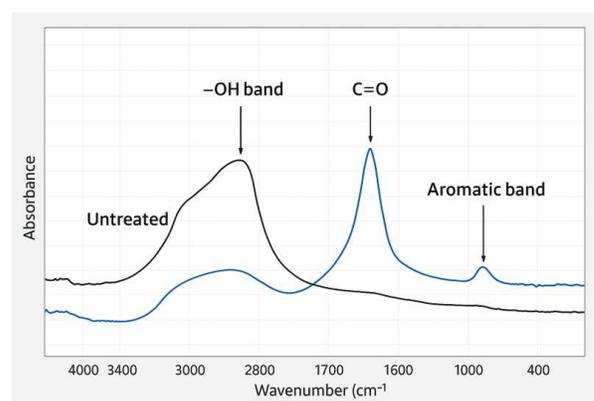

(b)

Fig. 6 : UV–Vis and FTIR spectral overlays, (a) UV–Vis spectra for control, PAW-10, PAW-15 showing bathochromic shifts; (b) FTIR spectra showing changes in –OH, C=O, aromatic bands post plasma treatment.

HPLC chromatograms confirmed these molecular modifications. Major peaks corresponding to catechin, eugenol, and azadirachtin exhibited shifts in retention time and relative peak area (Table – 2). A new minor peak at ≈ 9.3 min appeared in PAW-10 and persisted faintly in PAW-15, possibly representing a hydroxylated or nitrated derivative. The relative enrichment of such products is consistent with oxidative coupling pathways described by Abouelenein et al. (2022) and Pisheh et al. (2023).

Table – 2 : Representative HPLC peak-area variations of key bioactive compounds

| Compound | RT (min) | Area Control | Area (PAW-10) | Area (PAW-15) | Interpretation |
|---|---|---|---|---|---|
| Catechin | 5.2 | 1000 | 1150 | 1050 | Moderate activation |
| Eugenol | 7.8 | 800 | 920 | 860 | Derivative formation |
| Azadirachtin | 14.5 | 600 | 630 | 610 | Minor change |
| New peak | 9.3 | -- | 150 | 80 | Oxidized derivative |

*Applications and Implications*

The observed enhancement in antioxidant and antimicrobial properties suggests wide-ranging applications of PAW-modified herbal infusions.

Food Preservation: PAW-treated extracts can serve as natural preservatives in juices, dairy, or meat packaging, replacing synthetic stabilizers. The non-thermal plasma process minimizes nutrient loss and aligns with sustainable, additive-free food technologies (Patra et al., 2022; Zhang et al., 2023).

Biomedical Formulations: Enhanced antimicrobial and antioxidant effects make these extracts potential components of wound dressings, oral rinses, and nutraceuticals. Recent work shows plasma-generated RONS can synergize with bioactive phytochemicals for antimicrobial coatings and disinfection systems (Li et al., 2023; Chen et al., 2024).

Sustainable Processing: The plasma – herbal approach avoids chemical solvents, synthetic antioxidants, and harsh heating. By tailoring exposure time and plasma intensity, product functionality can be tuned for specific applications. Nevertheless, control of plasma dosage, long-term stability, and residual reactivity require further evaluation before industrial translation (Thirumdas et al., 2018).

Overall, this study substantiates the premise that plasma-activated water functions as a green modification tool that enhances phytochemical potency through mild, selective oxidation, offering a new frontier in natural-product processing and preservation.

## IV. CONCLUSIONS

This study demonstrates that plasma-activated water (PAW) can effectively modulate the structural and functional properties of neem (*Azadirachta indica*) and tulsi (*Ocimum sanctum*) infusions through controlled oxidative and nitrative interactions. The atmospheric-pressure gliding arc plasma employed here generated a balanced mixture of reactive oxygen and nitrogen species (RONS), notably OH•, NOγ, and excited $N_2$ species, creating a chemically active yet non-thermal environment. This composition proved essential for achieving beneficial molecular transformations without significant degradation of bioactives.

The results reveal that short-duration plasma exposure (10 minutes) enhanced both the total phenolic and flavonoid contents by approximately 10–15%, accompanied by a noticeable rise in antioxidant and antimicrobial capacities. These enhancements stem from mild oxidation and hydroxylation reactions that increase the electron-donating and radical-scavenging potential of polyphenolic compounds. The concurrent acidification of the medium promoted solubilization of bound phenolics, facilitating their release and transformation into more reactive forms. However, extending the plasma exposure to 15 minutes resulted in a partial decline of bioactivity, likely due to over-oxidation and breakdown of sensitive phytochemicals. This establishes an optimum plasma treatment window critical for maximizing functional improvements while preventing structural damage.

Spectroscopic and chromatographic analyses substantiated these transformations at the molecular level. UV–Vis spectroscopy showed bathochromic and hyperchromic shifts, indicating enhanced conjugation and delocalization of π-electrons within modified phenolic rings. FTIR spectra confirmed the formation of additional hydroxyl groups and altered hydrogen-bonding patterns, consistent with the incorporation of oxygen functionalities from RONS interactions. HPLC chromatograms further revealed distinct changes in retention time and peak areas for key compounds such as catechin, eugenol, and azadirachtin, including the emergence of new minor peaks suggestive of oxidized derivatives. Together, these findings validate that PAW induces selective oxidation and structural fine-tuning rather than indiscriminate degradation of phytochemicals.

The functional outcomes were equally significant. Antioxidant assays (DPPH and FRAP) indicated up to 20–25% improvement in scavenging and reducing capacities after PAW-10 treatment, supporting the hypothesis that moderate plasma exposure enhances redox efficiency. The antimicrobial assays demonstrated approximately two-fold reduction in MIC values against Escherichia coli and Staphylococcus aureus, suggesting a strong synergistic effect between residual RONS and plasma-modified phytochemicals. The improved antimicrobial efficacy likely arises from enhanced lipophilicity, improved permeability across bacterial membranes, and direct oxidative action of PAW-generated species.

These results highlight PAW as a promising green processing tool for herbal bioactive enhancement. The plasma–herbal synergy could revolutionize applications in food preservation, where PAW-treated extracts serve as natural antioxidants and antimicrobials for beverages, dairy, and meat systems, thereby minimizing chemical additive use. In biomedical and pharmaceutical formulations, PAW-treated infusions may contribute to advanced wound-healing materials, oral care rinses, and antioxidant nutraceuticals, leveraging their dual functionality. Moreover, sustainable processing benefits are clear, PAW eliminates the need for toxic solvents, conserves energy through non-thermal operation, and generates minimal environmental residues, aligning perfectly with the principles of green chemistry and circular bioeconomy.

However, certain limitations and future directions must be recognized. Long-term stability of PAW-modified compounds, potential residual reactive species, and large-scale reproducibility require further systematic evaluation. The influence of plasma parameters, discharge type, power input, gas composition, and treatment duration must be optimized to ensure consistent results across different herbal matrices. Additionally, mechanistic insights into specific reaction pathways leading to hydroxylation, nitration, or polymerization of phenolics should be elucidated through advanced mass spectrometry and molecular modeling.

Thus, the controlled application of plasma-activated water emerges as a viable and eco-innovative strategy to enhance the bioactivity of plant-derived infusions. By combining traditional phytochemical wisdom with cutting-edge plasma science, this approach opens a sustainable frontier for natural product preservation, functional food development, and biocompatible health formulations. The findings affirm that when appropriately optimized, PAW serves not merely as a sterilization medium but as a molecular activation platform transforming ordinary herbal extracts into potent, multifunctional agents for the bioeconomy of the future.